\documentclass[aps,prl,a4paper, nofootinbib, preprintnumbers,superscriptaddress,showpacs,showkeys,twocolumn]{revtex4-1}

\usepackage{amsmath}
\usepackage{amssymb}
\usepackage{bm}
\usepackage{graphicx}
\usepackage{color}
\usepackage[dvipsnames]{xcolor}
\usepackage[normalem]{ulem}  % \sout{old text} for strikeout
\newcommand{\be}{\begin{equation}}
\newcommand{\ee}{\end{equation}}
\newcommand{\ba}{\begin{eqnarray}}
\newcommand{\ea}{\end{eqnarray}}

\renewcommand\sout{\bgroup \color{blue} \ULdepth=-.5ex \ULset}

\begin{document}

\author{Maria Lucia Sambataro}
\email{sambataro@lns.infn.it}
\affiliation{Department of Physics and Astronomy “Ettore Majorana”, University of Catania, Via S. Sofia 64, I-95123 Catania, Italy}
\affiliation{INFN-Laboratori Nazionali del Sud, Via S. Sofia 62, I-95123 Catania, Italy}

\author{Salvatore Plumari}
\email{salvatore.plumari@dfa.unict.it}
\affiliation{Department of Physics and Astronomy “Ettore Majorana”, University of Catania, Via S. Sofia 64, I-95123 Catania, Italy}
\affiliation{INFN-Laboratori Nazionali del Sud, Via S. Sofia 62, I-95123 Catania, Italy}

\author{Santosh K. Das}
\email{santosh@iitgoa.ac.in}
\affiliation{School of Physical Sciences, Indian Institute of Technology Goa, Ponda-403401, Goa, India}

\author{Vincenzo Greco}
\email{greco@lns.infn.it}
\affiliation{Department of Physics and Astronomy “Ettore Majorana”, University of Catania, Via S. Sofia 64, I-95123 Catania, Italy}
\affiliation{INFN-Laboratori Nazionali del Sud, Via S. Sofia 62, I-95123 Catania, Italy}

\title{Probing the QGP through $p_T$-differential radial flow of heavy quarks}

%\vspace{10pt}
%\begin{indented}
%\item[]February 2016
%\end{indented}

\begin{abstract}
We introduce the $p_T$–differential radial flow $v_0(p_T)$ in the heavy-quark sector. Within an event-by-event Langevin framework, we show that this observable exhibits a strong sensitivity to the heavy quark–bulk interaction. It provides a powerful and novel tool to constrain the transport coefficients of heavy quarks in the QGP and, more generally, to assess the strength of the interaction of a Brownian particle in an expanding bulk medium. The results further indicate that heavy quarks exhibit collective behavior driven by the isotropic expansion of the QGP in heavy-ion collisions and, at low $p_T$, it offers a marked signature of the heavy quark hadronization mechanism. 
\end{abstract}

%\pacs{12.38.Aw,12.38.Mh}

%\keywords{Relativistic heavy-ion collisions, heavy quarks,  quark-gluon plasma, electromagnetic fields}

\maketitle

\textit {Introduction---} 
Strong evidence suggests that the Quark–Gluon Plasma (QGP)~\cite{Shuryak:2004cy, Jacak:2012dx}, a hot and dense deconfined state of quarks and gluons, has been observed in ultra-relativistic heavy-ion collisions (HIC) at Relativistic Heavy-ion Collider (RHIC) and the Large Hadron Collider (LHC). The evolution of the QGP is governed by strong pressure gradients that drive collective expansion, giving rise to various flow phenomena: anisotropic flows, originating from initial geometric asymmetries, and radial flow, resulting from the isotropic pressure-driven expansion of the medium. Over the years, differential measurements of particle azimuthal distributions have yielded strong evidence that the momentum distribution of final-state particles is a collective response to the initial pressure gradients. Radial flow, has so far been extracted from the slope of the transverse momentum spectra and characterized by a $p_T$-integrated radial-flow parameter. Thus, 
$p_T$-dependent information on radial flow has been missing, even though it plays a crucial role in the expansion of the QGP fireball. The recently proposed observable $v_0(p_T)$~\cite{Schenke:2020uqq,Parida:2024ckk} allows a $p_T$-differential investigation of radial flow, opening the opportunity to probe collective dynamics in the QGP arising from isotropic expansion, which has only recently become accessible. The $v_0(p_T)$ quantifies the correlation between the fraction of particles in a given $p_T$ bin and the mean transverse momentum of all the produced particle within a single event. Results from the ATLAS~\cite{ATLAS:2025ztg} and ALICE~\cite{ALICE:2025iud} Collaborations in Pb–Pb collisions at 5.02 TeV demonstrate the sensitivity of this observable to key properties of the medium, such as the bulk viscosity and the QCD equation of state \cite{Parida:2024ckk, ATLAS:2025ztg}. Furthermore, results from the ALICE Collaboration~\cite{ALICE:2025iud} (see also~\cite{Saha:2025nyu,Du:2025dpu}) indicate the mass-dependent behavior of $v_0(p_T)$ at low $p_T$ and reveal baryon–meson separation at intermediate $p_T$, a features poorly explored theoretically till now. 

Heavy quarks (HQs), produced in early stage of the collision due to their large masses, experience the entire space–time evolution of the QGP and, by retaining the memory of their interactions, serve as excellent probes of its transport properties \cite{Rapp:2018qla, Dong:2019unq,Das:2024vac,vanHees:2007me,Moore:2004tg,Gossiaux:2008jv,Uphoff:2012gb,Lang:2012cx,Alberico:2011zy,Das:2013kea,Das:2015ana,Song:2015sfa,Song:2015ykw,Cao:2016gvr,Cao:2018ews,Katz:2019qwv,Plumari:2019hzp,Beraudo:2021ont,Tang:2023tkm,Zhao:2024oma,Singh:2025duj}.
Measurements of heavy-quark nuclear suppression factors $R_{AA}$~\cite{ALICE:2021rxa} and elliptic flow $v_2$~\cite{STAR:2017kkh} at both RHIC and LHC provide constraints on the heavy-quark spatial diffusion coefficient $D_s$. Furthermore, the heavy meson-to-baryon ratio~\cite{ALICE:2018hbc,STAR:2019ank} serves as a probe of the hadronization mechanism, while the splitting in heavy-quark directed flow $v_1$ ~\cite{Das:2016cwd,Oliva:2020doe,Beraudo:2021ont} is considered a probe of the initial electromagnetic field.

In this letter, for the first time, we present a study of the $p_T$-differential radial flow of charm quarks in the QGP to probe the dynamics of  their in-medium interactions. For HQs the $v_0(p_T)$ is no longer driven directly by their contribution to the bulk viscosity and Equation of State of the bulk matter or the initial energy density fluctuations, because their contribution is largely subdominant; hence, the novel aspect is to exploit the event-by-event fluctuations in $p_T$ of a Brownian particle to determine the strength of its coupling to the expanding bulk medium, once the dynamics of the last is known.
We study the $v_0(p_T)$ of the heavy quarks in QGP within a event-by-event Langevin transport approach. In order to investigate the sensitivity of $v_0(p_T)$ to the heavy quark–bulk interaction, we explore a range of scenarios spanning from a weakly coupled regime, as described by perturbative QCD (pQCD), to a strongly interacting medium, as suggested by recent unquenched lattice QCD (lQCD) results \cite{Altenkort:2023eav, Altenkort:2023oms, HotQCD:2025fbd}.
The obtained $p_T$-differential radial flow is found to be highly sensitive to the underlying microscopic interaction dynamics, demonstrating that this observable, despite being independent of the mechanisms responsible for $v_2$ build-up, can be equally effective to provide quantitative constraints on the heavy-quark transport coefficient.\\
\textit {Event by event transport approach for charm and bulk dynamics---} We use an event-by-event (ebe) relativistic transport framework where we couple the ebe relativistic Boltzmann at a fixed shear-viscosity-to-entropy-density ratio $\eta/s$ with the Langevin equation \cite{Plumari:2012ep,Scardina:2012mik,Puglisi:2014sha,Scardina:2014gxa,Plumari:2015sia,Plumari:2015cfa,Scardina:2017ipo,Sambataro:2020pge,Plumari:2019gwq,Sun:2019gxg,Plumari:2019hzp,Sambataro:2025obe}.
The initial conditions of partons are described within a MonteCarlo Glauber Model, see ref.s \cite{Sun:2019gxg, Sambataro:2022sns,Sambataro:2023tlv,Sambataro:2025obe}. The soft partons are distributed in momentum space following an initial thermal equilibrium profile. We also consider a hard component consisting of minijets from initial binary pQCD collisions, with transverse momentum spectra at mid-rapidity taken from CUJET results for 5.02 TeV pp collisions \cite{Xu:2015bbz}.
Charm quarks are initialized in coordinate space using the binary collision profile $N_{\text{coll}}$ from the Monte Carlo Glauber model. The momentum distribution of charm quarks is computed from FONLL calculations which permits to reproduce the D-meson spectra in pp collisions after fragmentation \cite{Cacciari:2012ny, Scardina:2017ipo}.
In our approach, hadronization of charm quarks follows a hybrid model combining coalescence and fragmentation, detailed in Refs  \cite{Plumari:2017ntm, Minissale:2023dct, Minissale:2015zwa,Cao:2015hia,Gossiaux:2009mk,Oh:2009zj,Minissale:2024gxx}.\\
Regarding the particles dynamics, we describe the evolution of gluons (g) and light quarks (q) with the following equation:
\begin{equation}
p^{\mu}_{j} \partial_{\mu}f_{j}(x,p)= {\cal C}[f_{j},f_i](x,p) \, \, for \, \, i,j=g,q \label{eq:Boltzmann}
\end{equation}
where ${\cal{C}}[f_j, f_i](x,p)$ is the Boltzmann-like collision integral for elastic $2\rightarrow2$ scattering processes. 
Regarding the HQs dynamics, we describe their propagation in the QGP in a brownian motion by solving the following set of stochastic Langevin equations:
\begin{eqnarray}
 dx_j= \frac{p_j}{E}dt,  \hspace{10pt}\, \, \, \, dp_j= -\Gamma p_j dt + \sqrt{dt} C_{j,k} \rho_k \label{eq:y_lang}
\end{eqnarray}
where $\Gamma$ and $C_{j,k}$ govern the interaction between the HQ and the medium and they are directly connected to the \textit{drag} and \textit{diffusion} coefficients respectively \cite{Rapp:2008qc}. 
\begin{figure}[]\centering
\includegraphics[width=0.9\linewidth]{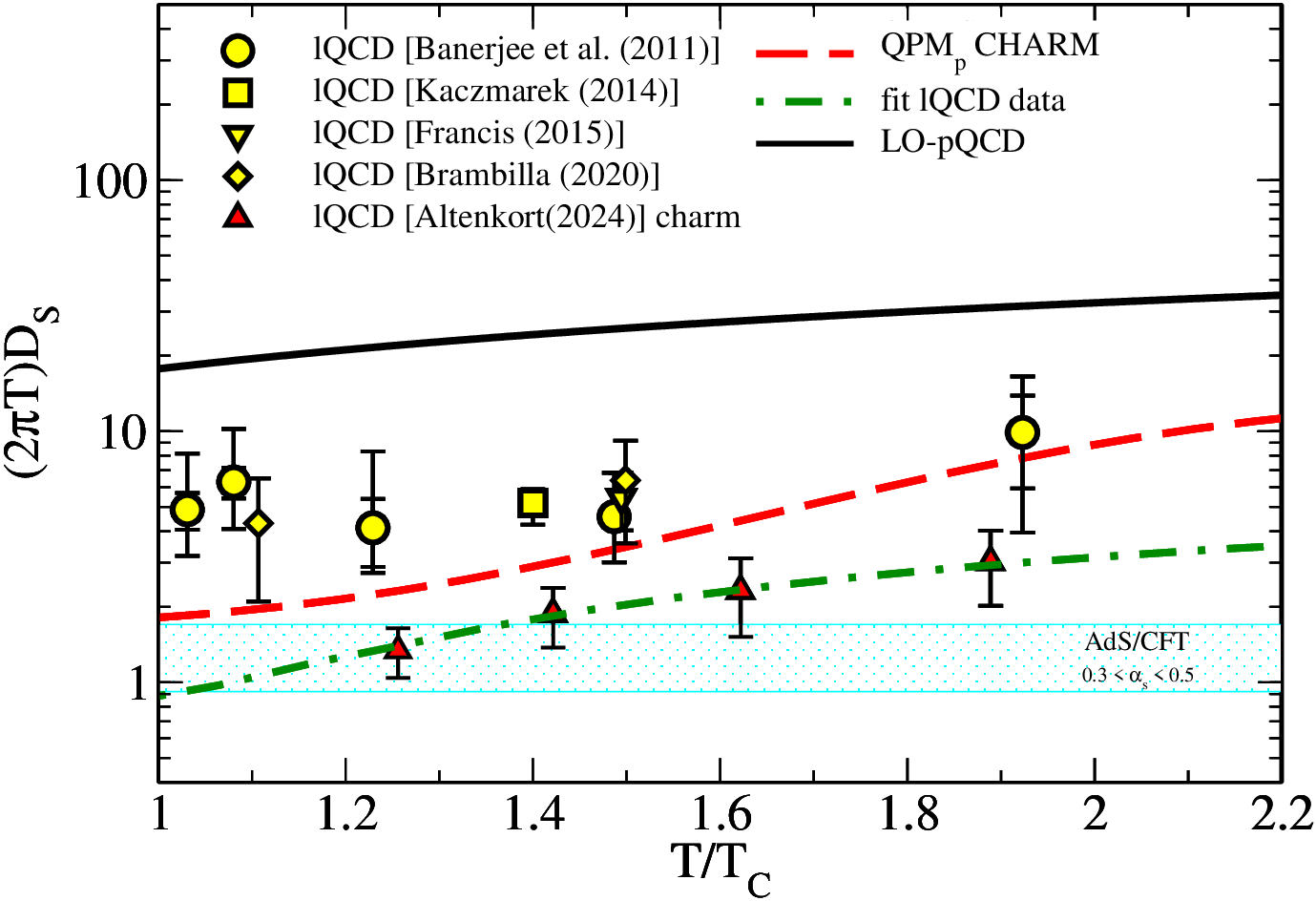}
\caption{Spatial diffusion coefficient $2\pi T D_s$ in $QPM_p$ and $T$-matrix approaches for charm quark compared to available lQCD data.}\label{fig1}
\end{figure}
We can express $C_{j,k} =\sqrt{2D_p(E)}\,\delta_{jk}$, where $D_p$ denotes the diffusion coefficient in momentum space. The fluctuation-dissipation theorem (FDT) then simplifies to $D_p(p) = A(p)E(p)T$. Notice that the drag coefficient $A$ and the momentum-space diffusion coefficient $D_p$ depend on $T(x)$, the bulk temperature obtained from the Boltzmann equation through a coarse-graining procedure. In the static limit $p \rightarrow 0$, we recall that $A(p \rightarrow 0) = \gamma$, and the momentum-space diffusion coefficient $D_p$ can be related to the spatial diffusion coefficient $D_s$ by the following relation:
\begin{eqnarray}\label{eq:Ds}
    D_s=\frac{T^2}{D_p}=\frac{T}{M_{HQ}\gamma}=\frac{T}{M_{HQ}}\tau_{th}.
\end{eqnarray}
In this letter, we consider three interaction scenarios between the charm quark and the bulk medium.
The first case corresponds to a weak interaction between the HQ and the QGP, considering drag and diffusion coefficients coming from leading order pQCD with a running coupling taken from ref. \cite{Kaczmarek:2005ui}. The corresponding $2\pi T D_s$ is shown in Fig.\ref{fig1} by black solid line. The second case corresponds to an extreme scenario of strong interaction, with a momentum independent $D_s$, green dot-dashed line in Fig. \ref{fig1}, fitted to the new unquenched lQCD data \cite{Altenkort:2023oms,Altenkort:2023eav,HotQCD:2025fbd}. These new lQCD data are significant smaller than the ones in quenched approximation (yellow symbols in Fig.\ref{fig1}), also suggesting a small thermalization time for charm quark $\tau_{th}\approx 1.5 \, fm/c$ at $T=1.2 \, T_c$ \cite{Altenkort:2023oms,Altenkort:2023eav,HotQCD:2025fbd}. In this second case both drag and diffusion coefficient are momentum independent with $A =T/M_{HQ} D_s$, resulting in a thermalization time constant in momentum. In the third case, we consider the drag and diffusion evaluated in a recently developed Quasi-Particle model, $QPM_p$, see red dashed line in Fig. ~\ref{fig1}. The $QPM_p$, which incorporates momentum-dependent parton masses consistent with QCD asymptotic freedom, successfully describes lattice QCD equation of state and susceptibilities of light, strange, and charm quarks \cite{Sambataro:2024mkr}. In the $QPM_p$, the effective vertex coupling $g(T)$ is derived from a fit to lQCD thermodynamics and is notably larger than the one predicted by pQCD approach, leading to a smaller $D_s$ especially as $T \rightarrow T_c$ (see details in Refs~\cite{Sambataro:2024mkr}). QPMp is also able to describe the $R_{AA}$ and $v_{2,3} (p_T)$ of D mesons for different centralities at top LHC energies \cite{Sambataro:2025obe}.\\
In this letter, we study the fluctuation of the $p_T$ spectrum of heavy hadrons associated with a fluctuation of the average heavy hadron transverse momentum $[p_T]$. 
Recently, in the light hadron sector a new observable called $v_0(p_T)$ has been introduced to measure this correlation, see \cite{Schenke:2020uqq}. It has been shown that the $v_0(p_T)$ in the light flavour sector depends on the bulk viscosity, also exhibiting a mass ordering going from pions to protons. The main goal of this work is to compute $v_0$ in the heavy-quark (HQ) sector and to analyze how this newly proposed observable may be sensitive to the interaction between HQ and the medium. 
Following the definition of $v_0(p_T)$ in~\cite{Schenke:2020uqq} we have:
\begin{equation}
  v_0(p_T) \equiv
  \frac{\langle \delta N(p_T)\,\delta p_T\rangle}
       {N_0(p_T)\,\sigma_{p_T}},
\end{equation}
which quantifies the event-by-event correlation between a fluctuation
$\delta N(p_T)$ of the single-particle spectrum at transverse momentum
$p_T$ and the fluctuation $\delta p_T$ of the mean transverse momentum
per particle.
For a given event, we denote by $N(p_T)$ the number of particles in a $p_T$ bin, i.e., the $p_T$ spectrum. For each event the spectrum $N(p_T)$ is decomposed as $N(p_T) = N_0(p_T) + \delta N(p_T)$, where $N_0(p_T)$ is the ensemble average. The event--wise mean transverse momentum is $[p_T] = N^{-1}\int_{p_T} p_T \, N(p_T) = \langle p_T \rangle + \delta p_T $, where $N = \int_{p_T} N(p_T)$ is the total multiplicity. 
The variance of these fluctuations is defined as $\sigma_{p_T}^2 \equiv \langle (\delta p_T)^2 \rangle$, where angular brackets represent averaging over events. By definitions, one can infer a characteristic behavior of $v_0(p_T)$ as a function of transverse momentum. Specifically, when the mean $p_T$ in a given event exceeds the ensemble average, the spectrum shifts toward higher $p_T$, leading to a positive $v_0(p_T)$ at high transverse momentum. This shift simultaneously induces a depletion of particles at low $p_T$, resulting in a negative $v_0(p_T)$ in that region.
Note that the formulas employed assume a constant particle multiplicity, i.e., $\delta N_{charm} = 0$. This condition is implemented in our approach for charm quarks, while the bulk medium is considered in its full realistic ebe flutuations in the selected centrality class, hence $\delta N_{bulk} \neq 0$. We notice that, instead, when defining and measuring $v_0(p_T)$ for light hadrons one keeps constant the multiplicities of the bulk, $\delta N_{bulk}=0$.\\

\textit{Results---}
\begin{figure}[]\centering
\includegraphics[width=0.9\linewidth]{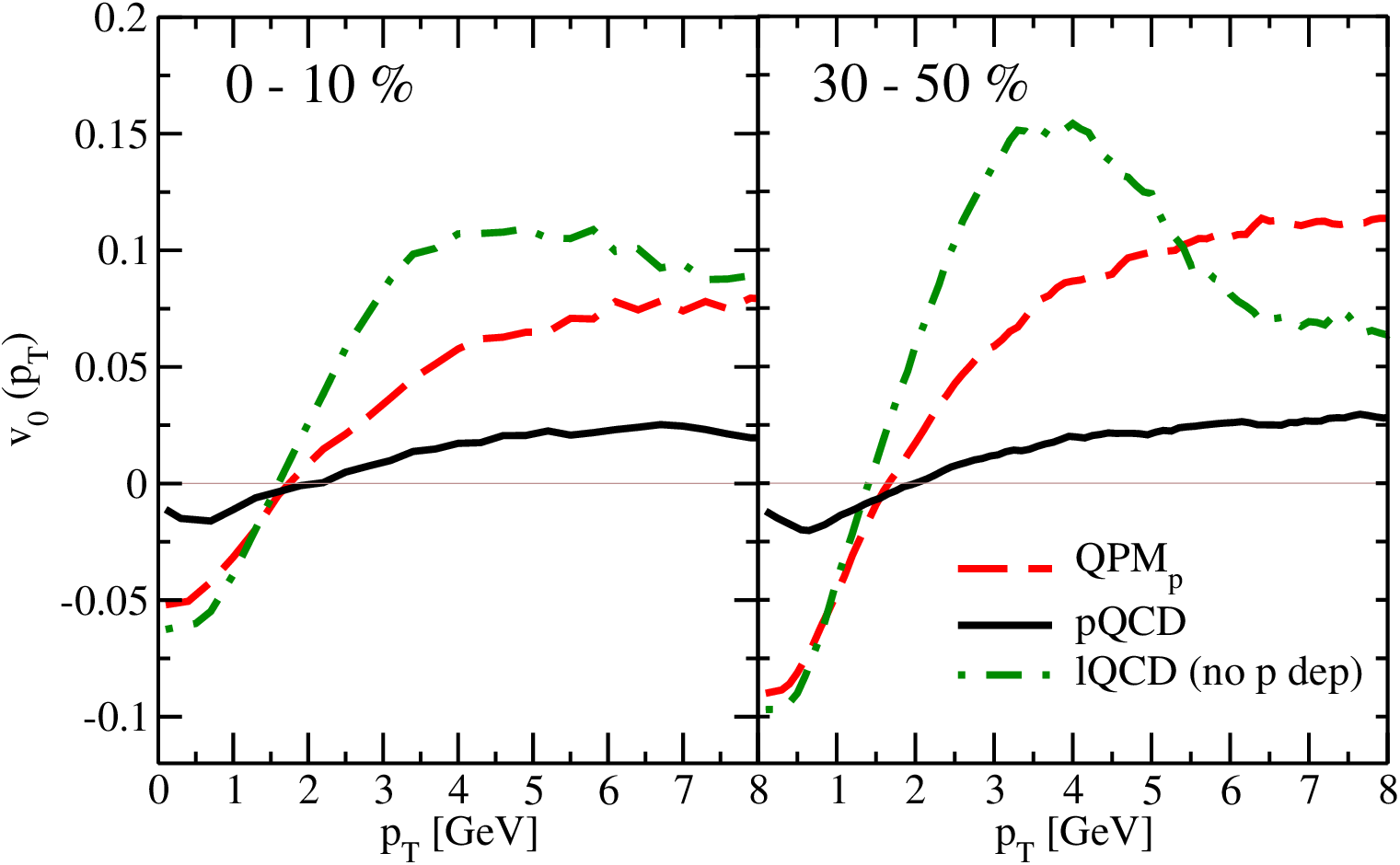}
\caption{$v_0(p_T)$ for charm quark at $0- 10 \, \%$ (left) and $30 -50 \, \%$ (right) in the different cases studied.}\label{fig2}
\end{figure}
First, we study the $v_0(p_T)$ at charm quark level for two centrality classes for Pb+Pb at $\sqrt{s}=5.02 \, TeV$  which is shown in Fig. \ref{fig2}. By construction, $v_0(p_T)$ is negative at low transverse momentum, as previously discussed, and changes sign around $p_T \approx M_c$. Specifically, we observe that in the $0–10 \, \%$ centrality class, the $v_0(p_T)$ obtained within the $QPM_p$ framework is approximately three times larger than in the weak interaction scenario at intermediate $p_T$. This enhancement becomes even more pronounced in the lQCD case, where $v_0(p_T)$ reaches values about four times higher than those predicted by pQCD which instead exhibits a flat trend with only mild dependence on transverse momentum $p_T$. 
As anticipated $v_0(p_T)$ is a sensitive measure of the coupling of charm quarks to the medium: for a weak interaction corresponding to a large diffusion coefficient $D_s$ (small drag of charm quark), as predicted by LO-pQCD, the medium ebe fluctuation are weakly transferred to the charm quarks implying a small value of the $v_0(p_T)$.
On the other hand, in a strongly interacting system such as that described by lQCD, the intense coupling leads to a more efficient transfer of bulk density fluctuations into the heavy-quark spectrum. As a result, the spectrum exhibits larger event-by-event deviations respect to the average, which manifest as a large $v_0$. Furthermore, our results indicate that, going from central to semi-peripheral collisions, $v_0(p_T)$ becomes increasingly sensitive to the differences in the spatial diffusion coefficient $D_s$ across the various models. 
In order to provide predictions directly comparable to an experimental observable, we consider a hybrid hadronization mechanism combining coalescence and fragmentation to  calculate the $v_0$ of D mesons and $\Lambda_c$ baryons.
\begin{figure*}[]
\centering
\includegraphics[width=0.9\columnwidth]
{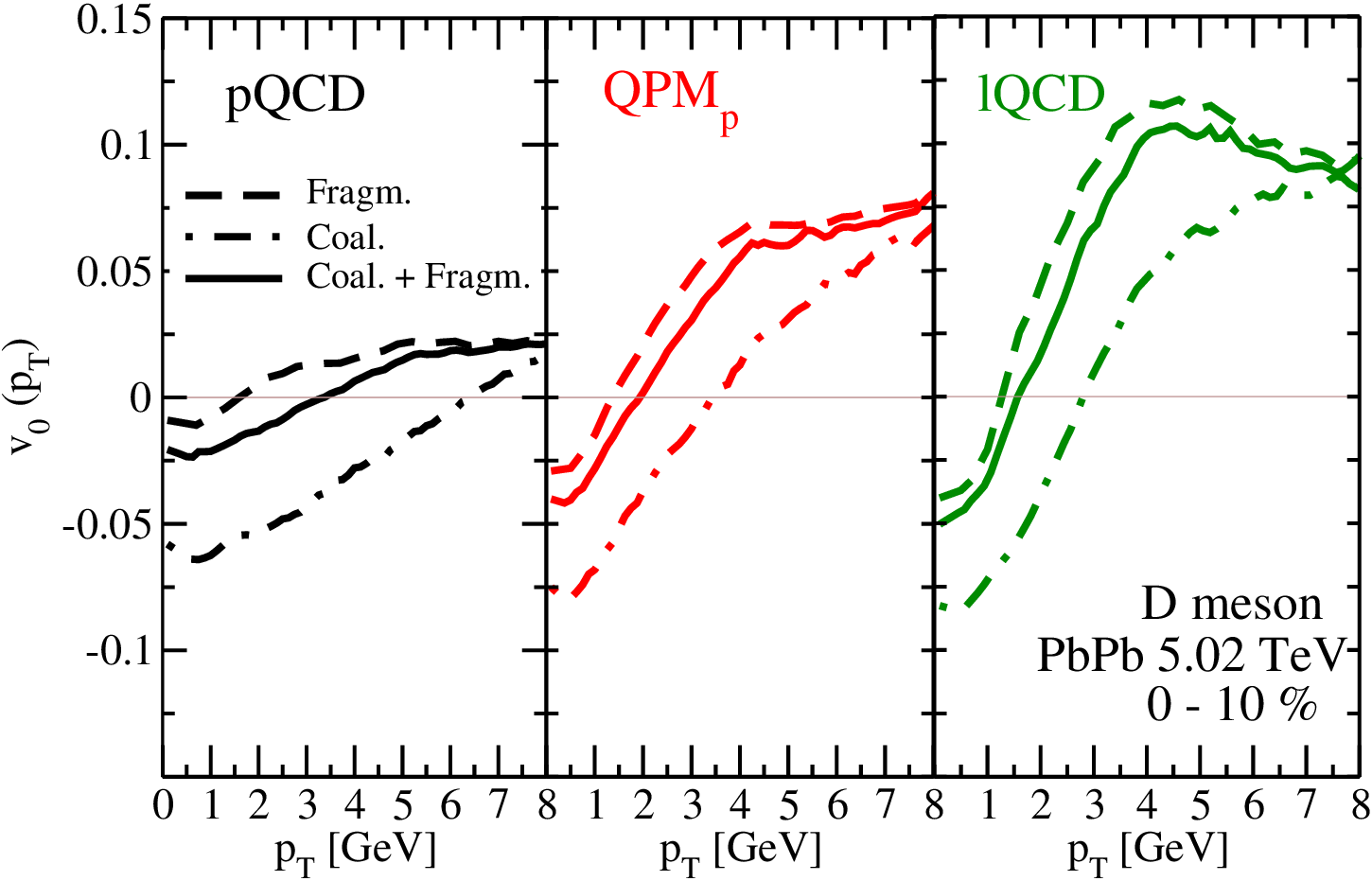}\quad \hspace{30pt}
\includegraphics[width=0.9\columnwidth]
{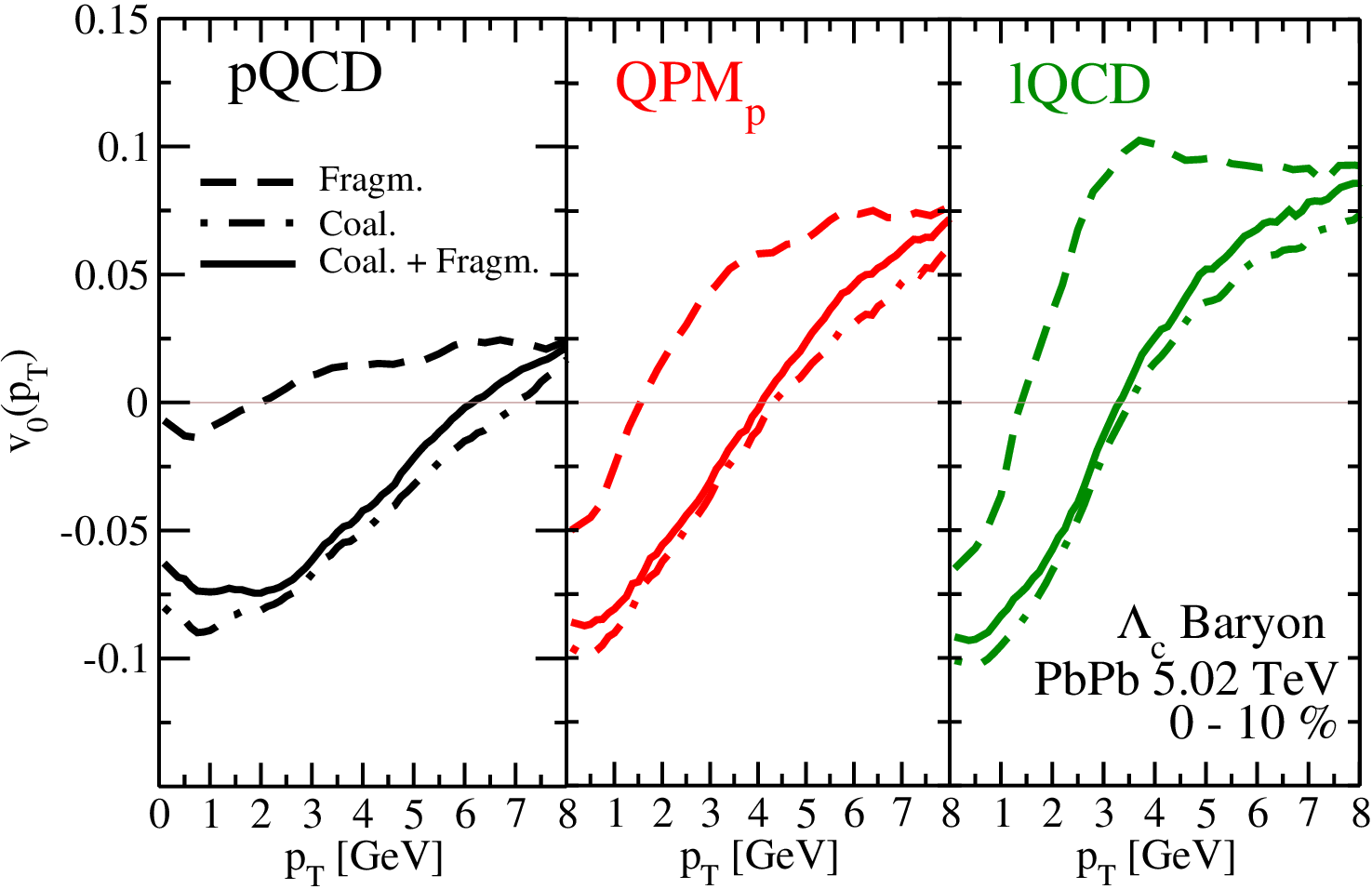}\quad
\caption{$v_0(p_T)$ at $0- 10 \, \%$ centrality class for D meson (left) and $\Lambda_c$ baryon (right) from fragmentation and coalescence for the different cases discussed.}
\label{fig3}
\end{figure*}
The key ingredients of the hadronization approach are presented in details in several papers \cite{Plumari:2017ntm, Minissale:2023dct, Minissale:2015zwa,Cao:2015hia,Gossiaux:2009mk,Oh:2009zj,Minissale:2024gxx}. It should be noted that an event-by-event coalescence-based hadronization mechanism still remains challenging at present, as it would require a fully dynamical, event-by-event coalescence approach. 
In order to get an approximate evaluation of the coalescence integral, we assume a $\delta$-like Wigner function, an approximation that carries the main feature of the coalescence process \cite{Greco:2004ex, Fries:2008hs, Fries:2025jfi}. Therefore, the hadron spectrum $f^H(p_T^H)$ in a single event can be written as the product of the spectra $f_i(x_ip_T^H)$ of its $N$ constituent quarks: $f^H(p_T^H)=C_H \prod_{i=1}^{N_q}f_i(x_ip_T^H)$.
The quark momentum can be written as $p_T^i=x_i p_T^H$ with $x_i$ the momentum fraction of quark respect to the hadron total momentum $p_T^H$, satisfying $\sum_{i=1}^N x_i =1.$. The spectrum of fluctuations take the form:
\begin{equation}
    \delta f^H(p_T^H)=C_H \prod_{i=1}^{N_q} f_i(x_ip_T^H)- \langle C_H \prod_{i=1}^{N_q} f_i(x_ip_T^H) \rangle
\end{equation}
which truncated at first order becomes:
\begin{equation}
    \delta f^H(p_T^H)=C_H \sum_{i=1}^{N_q} \delta f_i(x_ip_T^H)\prod_{j=1, j\neq i}^{N_q} \langle f_j(x_jp_T^H) \rangle
\end{equation}
We can write the hadron $p_T-$fluctuation as $\delta p_T^H=\sum_{i=1}^{N_q} \delta p_T^i$ and $\sigma^2_{p^H_T}\simeq \sum_{i=1}^{N_q} \sigma_{p_T^i}$. In our calculations, therefore assuming the off-diagonal terms $ \langle \delta f_i(x_ip_T^H) \, \delta p_{T_j} \rangle \simeq 0$ for $i\neq j$ the final hadron $v_0^H(p_T^H)$ can be written as:
\begin{equation}
    v_0^H(p_T^H)=\sum_{i=1}^{N_q} v_0(x_i p_T^H) \frac{\sigma_{p_T^i}}{\sqrt{\sum_{i=1}^N \sigma^2_{p_T^i}}}
\end{equation}
\begin{figure}[h!]\centering
\includegraphics[width=0.9\linewidth]{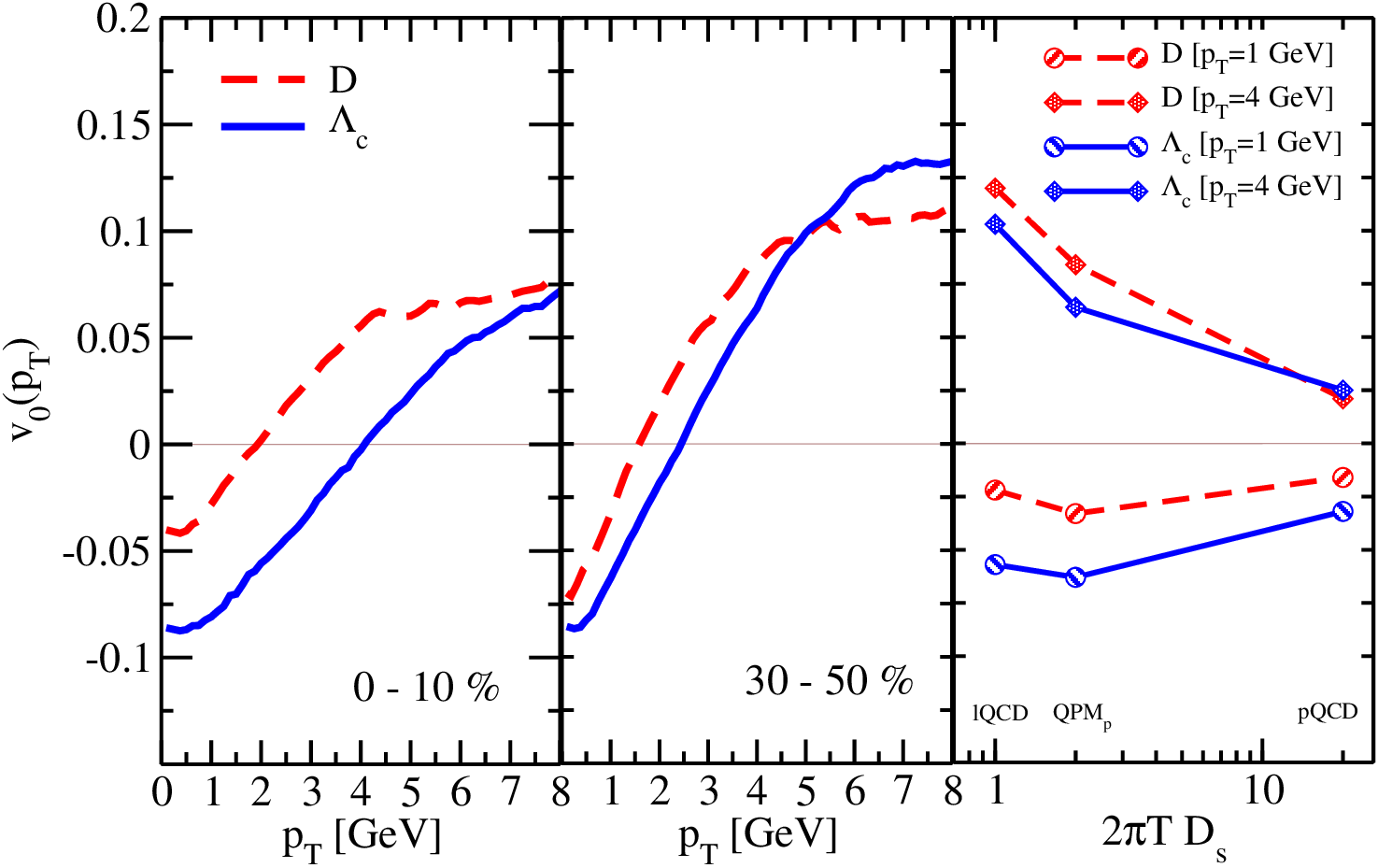}
\caption{$v_0(p_T)$ at $0- 10 \, \%$ (left) and $30-50 \, \%$ (middle) for D meson (red dashed line) and $\Lambda_c$ baryon (blue solid line) in $QPM_p$ from coalescence plus fragmentation. (right) D meson (red dashed line) and $\Lambda_c$ (blue solid line) $v_0(p_T)$ at $30 -50 \, \%$ centrality class for $p_T=1 \, GeV$ (circles) and $p_T=4 \, GeV$ (diamonds) as function of $2\pi T D_s$.}\label{fig4}
\end{figure}

Note that the $v_0$ of a hadron produced via coalescence results from a linear combination of the $v_0$ of its constituent quarks, in analogy with the sum rules derived for the $v_n$ coefficients \cite{Greco:2003mm,Kolb:2004gi}.
This implies that $\Lambda_c$ baryons are more affected by the $v_0$ of the bulk light quarks at low $p_T$ compared to D mesons. After the hadronization process, we observe that heavy flavour $v_0$ remains a sensitive observable to the interaction encoded in the $D_s$, as shown in Fig. \ref{fig3}, similarly to partonic level.
It is worth noting that the dominant contribution to the $v_0$ of D mesons arises from fragmentation, whereas for $\Lambda_c$ production, coalescence is expected to be the leading mechanism up to intermediate transverse momentum \cite{Plumari:2017ntm}. 
Regarding the $v_0$ of $\Lambda_c$, we observe a clear mass ordering at $p_T < 4 \,GeV$ as shown in the left and middle panels of Fig. \ref{fig4} where we plot the $v_0(p_T)$ at $0-10 \, \%$ and $30-50 \, \%$ for both D meson and $\Lambda_c$ baryon. Similarly to the pattern seen between pions and protons, charmed baryons exhibit a $v_0(p_T)$ that reaches zero-crossing at higher transverse momentum compared to D mesons. Notice also that the magnitude and centrality dependence of the heavy hadron $v_0(p_T)$ is quite comparable to that of light hadron measured at ATLAS~\cite{ATLAS:2025ztg} and ALICE~\cite{ALICE:2025iud} Collaborations for small values of $D_s$ (QPMp and lQCD). This indicates that the bulk fluctuations of $dN/dp_T$ in this case can be very efficiently transferred to the charm quark despite their large mass. The observed difference between heavy baryon and meson in $v_0(p_T)$ can provide insight into the hadronization process.
A main key result finding is that at low $p_T\simeq 1\,\rm GeV$ the $v_0$ is marginally dependent on the value of diffusion coefficient $D_s$ while it is marked the impact of the hadronization mechanism generating a difference of a factor of about two on the value of $v_0$ for $\Lambda_c$ versus $D$. This can be considered a fingerprint of coalescence for heavy baryons that recombine with two light quarks ate very low $p_T$ hence carrying a quite negative $v_0$ that significantly decreases the $v_0$ of $\Lambda_c$ wrt the D meson one.\\
\textit {Summary and Outlook---} We have introduced, for the first time, a new observable, $v_0(p_T)$, in the heavy-quark sector. We have pointed out that for a Brownian particle like HQ, the differential radial flow measures the coupling to the QGP bulk medium and the effectiveness of the response to its event-by-event fluctuations. This leads to a strong sensitivity to both the underlying heavy-quark transport coefficients (at intermediate $p_T$) and the hadronization mechanism (at low $p_T$).
Our study reveals that HQs in a strongly coupled regime should exhibit a collective behavior driven by the isotropic expansion of the QGP, with a magnitude comparable to that of light hadrons recently measured by the ALICE and ATLAS collaborations. 
This paves the way for introducing this new observable to achieve a quantitative constraint on the extraction of the spatial diffusion coefficient $D_s$ and a microscopic understanding of the underlying coupling to the bulk medium. 
Hence, its impact should also be investigated within a global Bayesian analysis.
We highlight the contrast between weakly coupled and strongly coupled HQ-bulk interactions with a difference of about a factor of 5 for $v_0(p_T)$ at intermediate $p_T$. 
After hadronization, $v_0(p_T)$ not only remains a sensitive observable of the interactions encoded at the partonic level,
but shows a specific pattern depending on the hadronization mechanism, particularly marked at low $p_T$. We have found that at low $p_T\simeq 1\,\rm GeV$ the $v_0(\Lambda _c) \sim 2\, v_0(D)$, while a pure fragmentation would give $v_0(\Lambda _c) \sim  v_0(D)$.
This makes the measurement of the $v_0$ of $\Lambda_c$ and $D$ a unique observable for revealing the hadronization mechanisms.
 
This letter represents a seminal work that opens several directions of investigation. Among them, we mention: the extension to pA and light AA systems down to OO collisions, which constitute a central focus of the HI-LHC program; the extension to bottom quarks, which, despite their larger mass, exhibit a similar $2\pi T D_s$ as charm quarks according to lQCD, thereby allowing the extraction of pure mass effects. Furthermore, such studies would provide new insights into bottom hadronization in AA collisions, which remains poorly understood. Another promising avenue involves exploring the relation between the momentum dependence of $v_0$ and that of the heavy-quark diffusion coefficient. Finally, we note that, more generally, $v_0(p_T)$ could be exploited to study the coupling of Brownian particles to an expanding mesoscopic systems.

\begin{acknowledgments} 
\textit{Acknowledgments---}
S.K.D. acknowledges Bedangadas Mohanty for useful discussion. V.G., S.P. and M.L.S. acknowledge the support of the PRIN2022 (Project code 2022SM5YAS) within Next Generation EU fundings and UniCT under PIACERI ‘Linea di intervento 1’ (M@uRHIC)PIACERI 2024-26.
\end{acknowledgments} 

%\bibliographystyle{JHEP.bst}
%\bibliography{biblio_HQ_v0}

%

\end{document}